\documentclass{elsart}
\usepackage{rotate}
\usepackage{epsfig}
\usepackage{graphics}
\usepackage{times}
\begin{document}

\begin{frontmatter}
\title{Integration and Conventional Systems at STAR}

\author[NSD]{H.S. Matis},
\author[BNL]{R.L. Brown},
\author[BNL]{W. Christie},
\author[LBLeng]{W.R. Edwards},
\author[LBLeng]{R. Jared},
\author[LBLeng]{B. Minor}, and
\author[LBLeng]{P. Salz}
\address[NSD]{Nuclear Science Division, Lawrence Berkeley National Laboratory, Berkeley, CA 94720 USA}
\address[BNL]{Physics Department, Brookhaven National Laboratory, Upton, NY 11973 USA}
\address[LBLeng]{Engineering Division, Lawrence Berkeley National Laboratory, Berkeley, CA 94720 USA}

\begin{abstract}
  
  At  the  beginning  of  the  design and  construction  of  the  STAR
  Detector,  the  collaboration  assigned  a team  of  physicists  and
  engineers the responsibility of coordinating the construction of the
  detector.  This group managed the general space assignments for each
  sub-system  and  coordinated  the  assembly  and  planning  for  the
  detector.  Furthermore, as  this group was the only  STAR group with
  the  responsibility  of  looking  at  the system  as  a  whole,  the
  collaboration assigned  it several tasks that  spanned the different
  sub-detectors.  These  items included grounding,  rack layout, cable
  distribution, electrical, power and water, and safety systems.  This
  paper describes these systems and their performance.

\end{abstract}
\end{frontmatter}

\section{Introduction}

The STAR  collaboration has built a detector  to study Nucleus-Nucleus
collisions  at the  Relativistic Heavy  Ion Collider  (RHIC)  which is
located at Brookhaven National Laboratory. The detector started taking
data  in the  year  2000 and  has already  published\cite{STAR1,STAR2}
several papers.   To achieve this,  many years of planning  and design
were needed.

When the project began, a group of physicists and engineers were given
the  responsibility  to  coordinate   the  activities  from  the  many
sub-systems.  This group  worked with the other STAR  groups so that a
coherently   planned  detector   was  constructed.    Each  individual
sub-system focused on the optimum performance of its design.  However,
this group,  which had a  global or overall system  level perspective,
assured that the individual plans worked well together.

As the sub-systems had several items in common, it was prudent for one
group to undertake the design of these common elements.  For instance,
the cooling  and powering of  each electronic rack is  essentially the
same for  each sub-system.  Similarly, as  the over all  cable plan is
common to each sub-system, it is  more efficient to build only one set
of cable  trays.  The integration  and conventional systems  group was
assigned  these  tasks in  addition  to  defining  the boundaries  and
interfaces between each sub-system.   What follows is a description of
this group's activities.

\section{Overall Layout of STAR}

\begin{figure}[ht]
%\vspace{75mm}
\centerline{\hbox{
\centering\psfig{figure=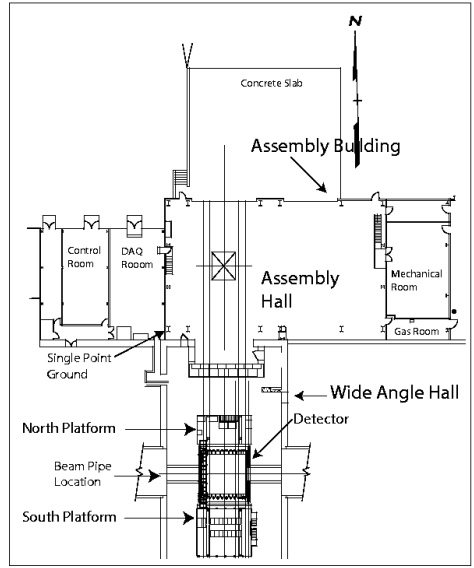,height=75mm}
}}

\caption{The layout of the STAR detector facility.  This diagram shows
the location of the various halls.   The detector is shown in the Wide
Angle Hall.}
\label{fig1}
\end{figure}

When the accelerator was  under construction, RHIC management assigned
the STAR detector\cite{IEEE} to the  Wide Angle Hall (WAH), located at
the 6 o'clock position in the accelerator ring.  A diagram of the full
STAR facility can  be found in Fig. \ref{fig1}.  As  the WAH was built
for  the  previous  accelerator,  Isabelle,  much of  our  design  was
constrained by  this existing structure.   A large addition,  which is
called the  Assembly Building, was needed to  accommodate the assembly
of the detector  and to house the magnet power  supplies and the water
and gas system.  The largest room in the Assembly Building, called the
Assembly Hall (AH), was used  to assemble the detector.  In this room,
the magnet and platforms were constructed. In a separate area, the TPC
was tested  and then  inserted into the  magnet.  The  detector easily
moves between the AH and WAH over two metal rails.

At the  center of the WAH, RHIC  beams collide in a  vacuum beam pipe,
which is located 4.31 m  above the floor.  The STAR detector surrounds
this beam  pipe and  provides the data  to analyze the  collisions.  A
platform is on each side  of the detector.  These structures provide a
location  for the  detector  electronics, pumps,  power supplies,  and
detector  utilities,  such  as  water,  power,  and  gas.   The  South
Platform,  which  is south  of  the  detector,  contains most  of  the
detector  electronics  while  the  North  Platform  holds  the  magnet
controls and power transformers.  It  is also a support for the magnet
and AC  power leads.  The North  and South Platforms  and the Detector
move as one unit.  Only a few utility connections such as power, water
and gas need to be disconnected before the detector can travel between
the WAH and AH.

As  any  item  in  the  WAH  can not  be  readily  accessed  when  the
accelerator operates,  we minimized the equipment  that resides there.
Nearly  all equipment that  must be  adjacent to  the detector  can be
operated remotely.  The electronics in the Data Acquisition Room (DAQ)
receive  trigger  and  digital  data  and also  communicate  with  the
detector through an optical Ethernet link.  This room is shielded from
the accelerator  beam by  concrete so that  it can be  accessed during
operation  of the beam.   Adjacent to  this room  is the  Control Room
where physicists monitor the data from the detector.

Next to the  Assembly Hall is a two-story  structure, which houses the
magnet power supply, gas and water systems.  The Assembly Hall is used
when the  detector needs extensive  servicing and provides  a facility
for testing sub-systems.

\section{Integration Drawings}

The STAR detector  has many layers.  As these  layers were constructed
at different  institutions, it was necessary to  coordinate the design
of each  element.  Therefore,  when such a  complex detector  is being
built, it  is necessary to  define guidelines so that  these detectors
can  come  together  cohesively.   As  the final  dimensions  of  each
sub-system are not known at the  start of the project, we defined what
was called an ``envelope''.

Each  sub-system  is  assigned  several  dimensions  in  a  series  of
drawings.\cite{envelope} It  is the responsibility  of each sub-system
manager to  assure that his/her sub-system fits  completely inside the
envelope.   To   ensure  that  there  are   no  interferences  between
sub-systems, space  is reserved for utilities,  signal cables, support
structures  and so  forth.  Often  a small  ``management  reserve'' is
inserted to account for future growth and changes.  This reserve could
be as  small as  1 to  2 cm. These  integration drawings  provided the
guidelines so that the whole detector would come together properly.

\begin{figure}[ht]

%\vspace{90mm}

\centerline{\hbox{
\centering\psfig{figure=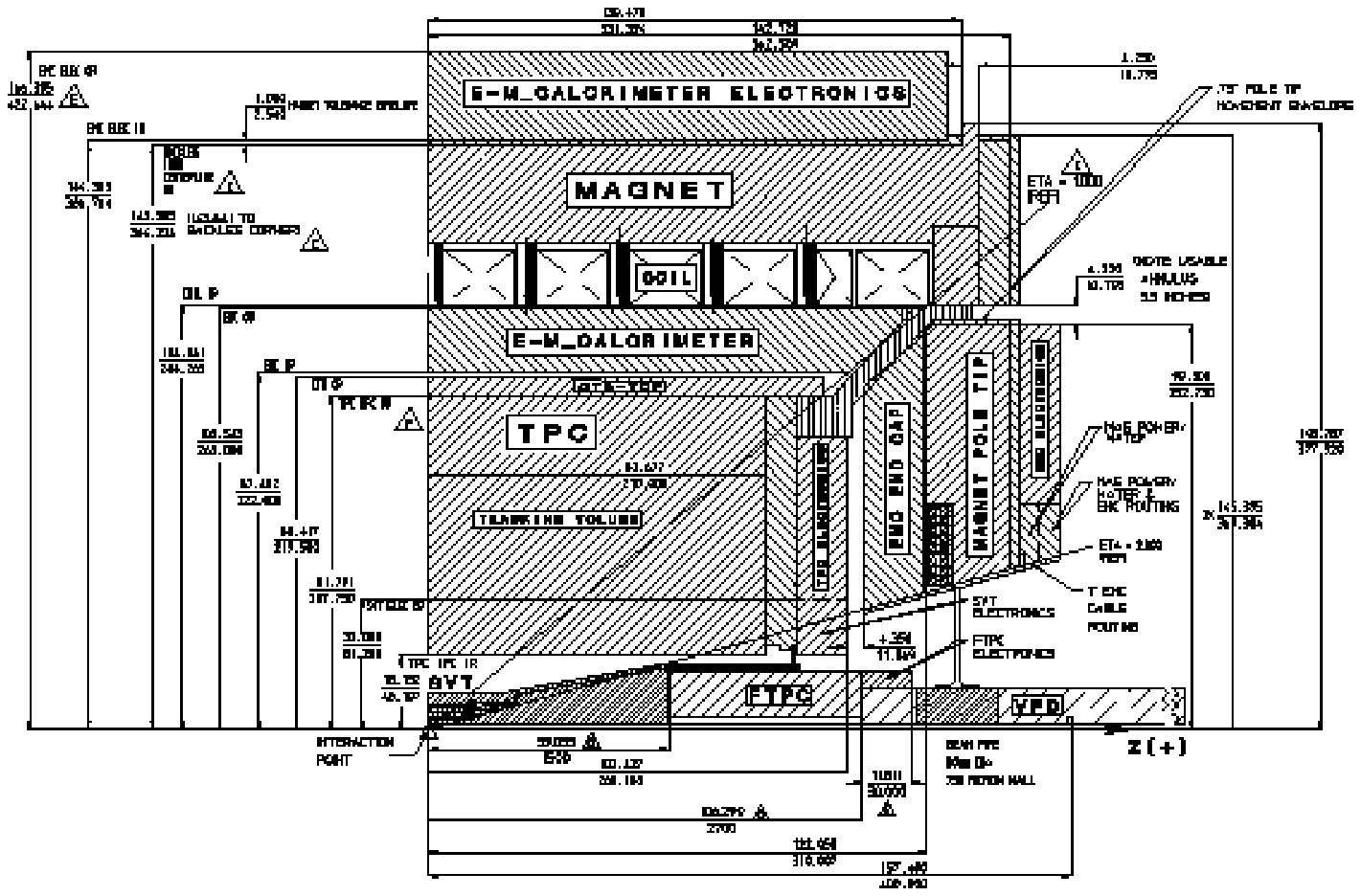,height=90mm}
}}
\caption{Quarter section of the  STAR detector. These regions describe
that  maximum  envelope  that   a  sub-system  can  fill.   The  upper
dimensions  are   in  inches  while   the  lower  dimensions   are  in
centimeters.}

\label{fig2}
\end{figure}

Figure  \ref{fig2} depicts a  quarter section  of the  detector.  This
drawing  shows  the  cylindrical  radius  and z  value  of  the  major
components  in the  detector.  This  figure is  just one  of  the many
series of  drawings needed to  define the dimensions of  the detector.
To  assure  that there  was  space  for  the future  additions,  these
subsystems were included in the envelope drawings.

\section{Safety -- Interlocks}

An interlock system has been designed\cite{safety1} and implemented to
protect  the STAR  detector  from damage.  The  STAR Interlock  system
consists of the interlocks for  specific STAR sub-systems as well as a
global system, which supervises those systems. All of these interlocks
utilize Allen-Bradley\cite{allen} Programmable Logic systems.

The  inputs to  this  system include  flammable  gas detectors,  smoke
detectors,  water  leak detectors,  and  the  status  of some  of  the
facility systems necessary to operate the detector (e.g. water cooling
systems). The actions that the interlock system can take automatically
include such things as removing AC power from either selected items or
the  entire detector.   It  can  also change  the  STAR flammable  gas
systems to purge modes with  nonflammable gases.  Upon sensing a water
leak, it can turn off water flow from portions of the detector, remove
the High Voltages for the electronics, and ramp down the STAR magnet.

Whenever  a fault  is  detected  by the  global  interlock system,  it
automatically puts the detector  in a safe state.  Simultaneously, the
system sends alarms both locally in the STAR complex as well as at the
RHIC main control room.

There  are  different  levels  of  response  depending  on  the  error
condition.   For  some conditions,  the  interlock  system provides  a
warning, leaving  time for human corrective action.   If the condition
is not  corrected after a  defined time, then automated  correction is
invoked.  For other conditions the interlock takes corrective measures
immediately.  Except for the relatively rare occasions when the entire
STAR detector is  being physically moved or when  the interlock system
is being modified, the system is always in operation.

\section{Grounding}

It is the  common experience of experiments to  find that the detector
in  an experiment  has significantly  more electrical  noise  then was
measured in  the laboratory. A  frequent culprit to noise  problems is
poorly planned  electrical distribution.  There may  be many different
and ill-defined electrical paths, resulting in current loops, which in
turn produce spurious electrical signals.

To reduce electrical noise pickup requires a single point ground and a
well-planned  distribution of  electrical power.\cite{ground1,ground2}
Care must  be taken  to avoid coupling  from noisy  electrical devices
such  as motors,  pumps  and  switching power  supplies.  As only  the
physical building existed before the STAR detector was constructed, we
were able to design an electrical system with few prior constraints.

While  it  is  possible to  make  sure  that  the  detector has  a  DC
resistance greater than 100 M$\Omega$  from any sources other than the
single point ground, it is  very difficult to completely eliminate the
capacitive coupling.  Therefore, we had to make an electrical model of
the STAR detector.  To gauge how much capacitance would be acceptable,
we examined the effect of it on the most sensitive element of the base
line STAR  detector -- the TPC digitizing  electronics.  For instance,
the capacitance  of a TPC  pad is 30  pf with about 1000  electron RMS
noise.  It was necessary to control the STAR electrical environment so
that this noise was not increased significantly on the TPC pads.

\begin{figure}
%\vspace{64mm}

\centerline{\hbox{
\centering\psfig{figure=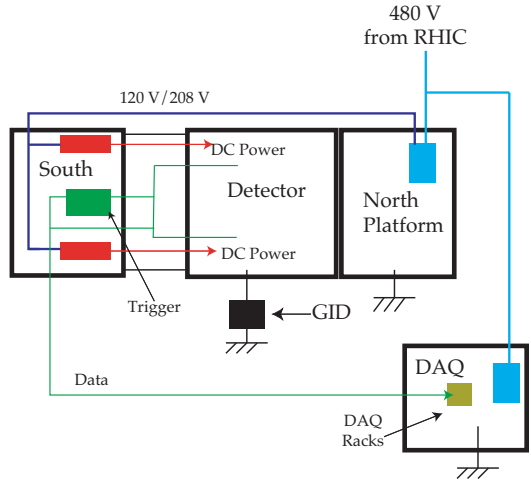,height=64mm}
}}

\caption{ This schematic  shows the clean power flow  and grounding of
the STAR detector.  Both the  South Platform and the Detector have the
same ground.  The location of  the Ground Integrity Detector  (GID) is
indicated just below  the center of this figure.   The 208/120 V power
is isolated from the ground of the North Platform.}

\label{fig3}
\end{figure}

We  devised   a  grounding  plan\cite{grndNote}   that  separated  the
electrical components from each  other, reduced stray capacitance, and
split the  electrical power distribution.  The schematic  for the plan
is shown in Fig. \ref{fig3}.  We mandated that the detector should use
optical connections  when possible, that the  detector be electrically
isolated  in  DC resistance,  and  that  the  capacitance between  the
detector and its surroundings be held to less than 5 nf. This value of
capacitance was  selected so that anticipated noise  sources would not
appreciatively affect the TPC front-end electronics up to 100 kHz.

A column in the Assembly Hall was selected at the earth contact of the
experiment.  A  total of  eight 4/0 cables  ground the magnet  to this
column.  These   cables  provide  the  single  point   ground  to  the
detector. We selected this column because  is about 30 m away from the
beam  pipe, and consequently  beam induced  electrical noise  would be
attenuated before it reached the  column.  It was also in a convenient
place to ground the  detector when it is either in the  WAH or AH.  It
is also  adjacent to the  DAQ Room so  that both the detector  and its
electronics have the same ground.

The trigger  and power  supplies to the  detector reside on  the South
Platform.   These items  communicate  with the  DAQ  room via  optical
cables.  The  North Platform, which is electrically  isolated from the
detector,  holds items  such  as the  480V  transformers and  magnetic
monitoring  equipment.   We  ordered transformers  with  electrostatic
shields so that electrical noise from the power grid did not enter our
power distribution system.

\section{Facility AC Power}

The 6 o'clock facilities for the Detector have AC power feeds from two
separate substations.  The first  substation is a  13.8 kVA  feed that
supplies 480 V transformer power to detector power panel A1 located in
the  DAQ room.  This  panel provides  power  distribution to  detector
electronics through three shunt trip breakers A11, A12, and A13. Shunt
trip breakers are used as  required to interlock all detector AC power
with the STAR Global Interlocks System (SGIS).

Circuit A11  provides ``Clean Power''  to 480 V  isolation transformer
located  on  the  North  Platform. From  these  isolation  transformer
208/120 V  clean power is distributed to  individual electronics racks
on the  South platform  through rack row  breaker panels.   The second
circuit, A12,  supplies ``Dirty  Power'' to 480  V transformer  on the
North Platform  for 208/120 V  power distribution. This power  is used
strictly  on the  North Platform  to  provide lighting  and power  for
detector  subsystem rotary  equipment that  will generate  line noise.
The last circuit, A13, provides DAQ  and Control Room Power to a 480 V
isolation  transformer located in  the DAQ  room. From  this isolation
transformer 208/120  V clean power  is distributed to  DAQ electronics
racks and to Control Room computer systems.

Shunt  trip  breakers  interlocked  to  SGIS  are  assigned  to  those
electronic racks  that are temperature sensitive  and require Modified
Chilled Water (MCW)  cooling and leak detection.  On  each rack, there
are P2  rack breaker  panels which distribute  208/120 V power  to the
electronics.

The  second AC  power feed  comes  from a  13.8 kVA  overhead line  to
substation-8  located on  the East  Side  of the  Facility. This  feed
provides power  to the  facilities buildings for  conventional systems
use and  also to power the STAR  Magnet DC power supplies.  There is a
direct 13.8 kVA feed to the Main Magnet switch gear that in turn feeds
the  Main  Magnet transformers  and  rectifier,  which  can provide  a
maximum of 5000  A @ 600 VDC. There  is also a 480 V feed  to the four
other Magnet Trim transformers and rectifiers used to control magnetic
field quality in the detector.

\section{Rack Design}

To preserve the  life of the electronics, it is  important to have the
electronics operate close to room temperature. It is widely known that
as   the  temperature  of   electronics  increases,   the  reliability
decreases.   Because  of  the  extremely  limited space  on  the  STAR
platform,  we had  to pack  the  electronics as  densely as  possible.
Since we did not want a situation where there was excessive heat build
up inside the rack, we  decided to put water-cooled heat exchangers in
each crate.

\begin{figure}
%\vspace{78mm}
\centerline{\hbox{
\centering\psfig{figure=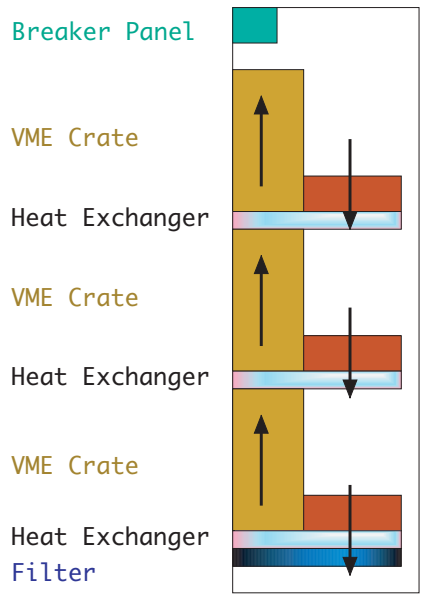,height=78mm}
}}

\caption{A side view of the air flow  in a rack.  Each VME crate has a
fan in the front  of the crate, which forces the air  up.  At the rear
of the VME crate, the power  supply blows the air downward.  On top of
the rack is a P2 breaker panel.}

\label{fig4}
\end{figure}

Figure  \ref{fig4} shows  a typical  arrangement for  a rack  with VME
crates.     The   heat    exchangers   are    modeled    after   those
designed\cite{rack} for the D0 experiment.  The fans inside the 6U and
9U sections of the VME crate blow the air up.  The fans in the back of
the crate move  the air down to facilitate the  airflow. A filter held
by a  simple metal  frame traps  dust inside the  rack.  A  P2 breaker
panel, which  distributes electrical power,  is located on top  of the
rack.

Racks are  arranged so  that two 19"  racks occupy one  volume.  Blank
panels  fill  all unoccupied  outside  surfaces.   This minimizes  the
amount  of air  exchanged in  the  rack and  consequently reduces  the
amount of  dust.  Furthermore, keeping  the air inside the  rack makes
smoke detection much more efficient.

\section{Ground Integrity Detector}

As the construction of a detector is very complicated and done by many
people, we had a significant  concern that the grounding plan would be
inadvertently  violated.   Therefore   we  built  a  Ground  Integrity
Detector (GID), a device which checks to see if there is an electronic
path  to ground  other than  the  intended single  point ground.   The
device sends an AC 400 Hz  signal current through the ground leads and
then checks whether all current is  returned.  It enables us to have a
dynamic determination of the integrity of the ground.

\begin{figure}
%\vspace{50mm}

\centerline{\hbox{
\centering\psfig{figure=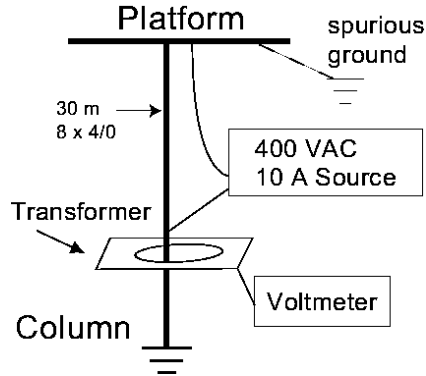,height=50mm}
}}

\caption{This schematic shows the supervisory current connection.  The
item, labeled ``spurious ground'',  indicates a path that would divert
current that passes through the transformer.}

\label{fig5}
\end{figure}

The physical layout, which is depicted in Fig. \ref{fig5}, consists of
a  grounded stanchion with  a welded  on boss  for mounting  a current
transformer.  The boss  has a flange to secure the  4/0 cables used to
ground the  platform.  The current transformer detects  current in the
boss. A  supervisory current (400 Hz)  is sent through  the 4/0 cables
from the cable  flange to the platform.  The  transformer does not see
this current.   If an extraneous path  from the platform  to ground is
accidentally created  then that current must pass  through the current
transformer.  The voltmeter will detect this current.

The current  transformer output consists of a  voltage proportional to
the  current  through  the  mounting  boss.  To  guard  against  false
readings  due   to  60  Hz   leakage  current,  the  signal   must  be
filtered. The electronics  consists of two 400 Hz  bandpass filters in
tandem preceded  by a  300 Hz  high pass circuit.   The output  of the
second band pass filter is fed to an active rectifier chip and then to
a  Simpson  controller.   The  controller  provides  relative  voltage
(proportional to extraneous  400 Hz current) as well  as preset visual
and audible alarm functions.

The eight  4/0 cables are  used as a  single point ground and  are the
safety ground for  the electronic platform The DC  and 60 Hz impedance
of the cable array is  approximately 0.7 m$\Omega$. However the 400 Hz
impedance is about  1.7 m$\Omega$.  If a droplight  that was connected
to another power  source were inadvertently hung on  the platform, the
ground  wire  in the  cord  would provide  a  leakage  current to  the
supervisory  current.   The  magnitude   of  the  leakage  current  is
dependent  on many  factors,  but  is typically  300  mA.  This  value
provides  a  current transformer  output  of  350  mV at  400Hz.   The
detector has a sensitivity of 100 mA = 0.168 $\Omega$ or about 30 m of
number 12 wire.

\section{Fiber and Cable Plan}

The STAR detector differs from  previous collider detectors as much of
the digitization  is on the  detector.  This design requires  that low
voltage power  cables are  the largest volume  of cables  entering the
detector.  Because of  grounding and the desire to  reduce the size of
the  power supplies, the  low voltage  power supplies  locally provide
power to  the detector  from the platform.   These cables  were routed
from the  platform to radial cable  trays on the magnet  face.  All of
the  data, except  for the  trigger detector,  goes directly  from the
inner section of the detector and then is routed to the south platform
and then  to a very simple group  of movable supports.  At  the end of
the supports,  the cables  pass through a  concrete tunnel to  the DAQ
room.

The readout between  the detector and Data Acquisition  Room is almost
entirely made up of fiber optical cable.  Because adding connectors in
these cables attenuates the signal and reduces the reliability, it was
decided that the detector readout  chain must be left intact and never
broken.  Consequently,  we designed  a  flexible overhead  ``festoon''
system for  transporting the fiber cable  bundle from one  hall to the
other.  This  system required the addition of  approximately 35 meters
of fiber to the total cable run of approximately 100 m.

The festoon easily  moves as the detector travels from  the WAH to the
AH.  As a  result there is no need to disconnect  the data cables when
the detector is  moved.  As the South Platform is  rigidly tied to the
magnet,  there was no  need to  disconnect the  power cables  from the
detector.

The distance  via the cable trays  between the South  Platform and the
detector vary by as  much as 11 m because the trays  are routed on the
radius of  the Magnet.  If the  cables were cut to  length, then there
would be  different timing for  signal cables and  different resistive
losses for power cables. Because  we desired that the cable effects be
similar, we decided  that all similar types of cable  should be cut to
the  same length.   This decision  resulted in  us  having significant
amounts of excess  cable.  These surplus lengths were  stored in cable
trays  that  run  under the  second  and  third  floors of  the  South
Platform, and under each floor  of the South Platform. The floors were
designed so that  there was easy access to the  cable trays. A typical
length for a cable from the South Platform to the detector is 35 m.

\section{Water Systems}

The STAR Facilities have as part of their conventional systems a stand
alone  water  system.  The  Detector, in  full  operation,  dissipates
approximately  4 MW of  electrical power  through various  closed loop
water systems with heat exchange to an open air cooling tower.

The magnet closed loop system is a Low Conductivity Water (LCW) system
dissipating  3.2 MW using  1100 gpm  @ 200  psi while  maintaining the
aluminum  coil  assemblies  at a  mean  temperature  of  29 C  with  a
$\Delta$T of 11 C. Because the conductor is aluminum, all materials in
this  system are  restricted to  aluminum and  stainless steel  with a
water  resistivity  greater  than  5 M$\Omega$-cm.  This  system  goes
through two heat exchanges. the first time with the cooling tower as a
pre-cooler,  and the  second with  a  chiller system  used during  the
warmer months of the year.

Detector electronics, represents approximately 160 kW of power that is
dissipated through  the Modified Chilled Water (MCW)  system. This LCW
closed loop water  system provides 16 C water using 360  gpm @ 150 psi
through a  heat exchange  with a chiller  system. There  are secondary
closed loop  systems that  operate off a  heat exchanger with  the MCW
system for cooling TPC, SVT, TOFp, and FTPC detector subsystems.

The  Magnet DC  power supplies  and water-cooled  electrical  buss are
another closed loop water system that dissipates 340 kW through a heat
exchange  with the  cooling  tower. As  mention  previously, both  the
Magnet and MCW  water system dissipate a large  portion of their heat,
in the warmer months of the year, through a heat exchange with a water
chiller system.  These 2  - 200 ton  capacity chillers are  capable of
dissipating the 1220 kW load from both the magnet and MCW systems in a
heat exchange with the cooling tower.

Programmable Logic  Controllers (PLC)  control all water  systems with
alarm and  interlock capability.  The system can  be monitored  in the
STAR  Control Room  through an  interactive schematic  display  of the
system parameters of each water system.

\section{Beam Pipe}

\begin{figure}
%\vspace{45mm}

\centerline{\hbox{
\centering\psfig{figure=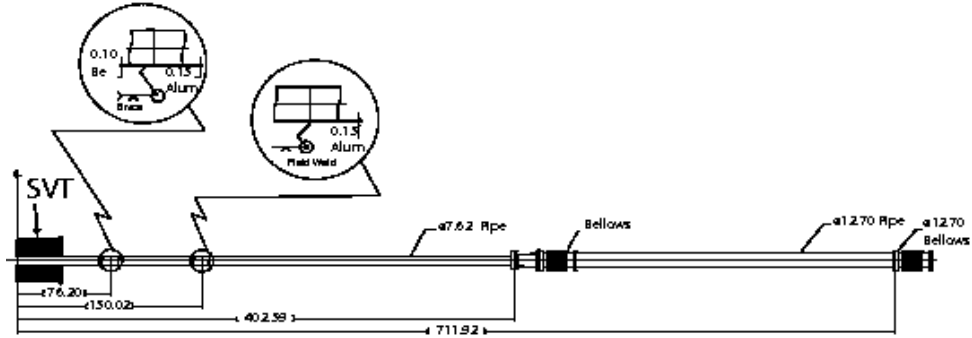,height=45mm}
}}

\caption{Half section of the STAR  beam pipe.  The envelope of the SVT
is  drawn to  show where  the  transition from  beryllium to  aluminum
occurs.  All units  are in cm.  The beam pipe is  supported by the SVT
and by a magnet adjacent to the end bellows.}

\label{fig6}
\end{figure}

The  beampipe  assembly  for  STAR   spans  the  16  m  width  of  the
experimental  hall. Located  at  the edges  of  the hall  are the  two
interaction  region vacuum pumps.   The beampipe  near these  pumps is
12.7 cm in diameter; but it necks down to only 7.62 cm at 4 m from the
interaction  point (IP).  A layout  of the  beam pipe  can be  seen in
Fig. \ref{fig6}.

The goal for  the beampipe and pumping system design  is to maintain a
vacuum in  the $10^{-9}$  torr range  at the IP  in order  to minimize
beam-gas  interactions.  One of  the more  challenging aspects  of the
design was to maintain this  level of vacuum while minimizing the mass
of  the beam  tube and  supports within  the constraints  of  a nearly
hermetic  collider   detector.   As   is  typical  of   most  collider
experiments,  there  are many  detector  elements  that cluster  close
(within a  few cm) to the beampipe  and extend the full  length of the
detector.   Additionally,  these subdetectors  must  be installed  and
removed  for maintenance  along the  beam axis.   This means  that the
opportunities  for supporting the  beampipe are  few and  far between.
The last structural item of significant  mass is located over 4 m from
the IP.

To  minimize  the  material near  the  IP,  we  use a  cone  structure
fabricated  from a  carbon fiber-epoxy/Nomex  honeycomb  material that
holds the Silicon  Vertex Tracker (SVT) to support  the beam pipe.  At
54.8 cm from  the IP, a spider arrangement of  4 aluminum radial wires
(3.18 mm  diameter), 4  G-10 insulating rollers  (7.19 mm  diameter by
25.4 mm long) and  a disk (128 mm O.D. by 89 mm  I.D. by 6.4 mm thick)
fixes the beampipe to the SVT cone.  The SVT cone then attaches to the
TPC wheel, which is mounted to the magnet.

While  the beampipe  outside the  STAR  detector is  1.5 mm  Stainless
Steel, as  it transitions to the  smaller 7.62 cm  diameter within the
detector it becomes 1.24 mm thick  Aluminum, then at 76 cm from the IP
becomes 1 mm wall Beryllium.

This design utilizes a single standard length of rolled Beryllium tube
at the heart  of the detector.  To reduce the cost,  the length of the
Beryllium  is limited and  a thin  Aluminum section  is welded  to it.
This economical choice only  minimally impacts the number of secondary
or stray particle  hits in the SVT and TPC.   The Beryllium is cleaned
and passivated to  minimize oxidation and Be dust so  that it safer to
handle.   It was  joined  via  braising to  a  short 6061-T6  aluminum
transition  and  welded  to  an Aluminum/Stainless  Steel  ``Conflat''
flange by  the manufacturer.  After arriving at  the STAR experimental
hall, the end  was cut and additional lengths  of aluminum tubing were
field welded in place to make up the full-length 8.08 m pipe.

Many considerations come  into play in determining the  beam tube wall
thickness.   The  thickness  is  a pseudo-optimization  of  acceptable
safety  factor from ``critical  buckling pressure''  calculations, the
allowable  mid-span/mid-support  deflections as  well  as standard  or
available tube wall  thickness and the necessity to  be able to handle
and transport  the pipe.  The  analyses included the  reduced material
strength  and  stiffness  at  bakeout temperatures  ($\sim$150  C  for
Aluminum),  the imperfections  in  tube ``roundness'',  and the  extra
weight  and space  requirements of  bakeout blankets.   We  selected a
safety  factor between  3 and  4.   This value  produces a  relatively
robust  design  that should  survive  the  occasional maintenance  and
upgrade  situations   while  minimizing  both   the  multiple  coulomb
scattering for  particles that are  produced inside the beam  pipe and
the secondary particles that are created by the beam pipe.

\section{Mechanical Movement of the Detector}

The  STAR Detector must  move 33  m between  the Assembly  Hall, where
maintenance and  upgrades occur, to  the Wide Angle  Hall.  Mechanical
movement  of the  1100  Tonne detector  is accomplished  hydraulically
using the  inchworm principle.  A set  of thick rail  plates, on which
the 4 -  500 ton Hillman Rollers ride, are imbedded  in the hall floor
perpendicular  to  the detector  (beam)  axis.   These  plates have  a
repeated hole  pattern for  bolting the base  of the  30-Ton hydraulic
actuators used to move the detector in approximately 1.5 m increments.
These  actuators both push  (extend) and  pull (retract)  with similar
load carrying capability.

The detector  was designed and  constructed with the  cylindrical axis
1.3  cm below beam  height, when  resting on  the Hillman  Rollers and
floor rail  plates. When  the detector  is moved into  the WAH,  it is
raised  vertically  to beam  height  and  supported  using 4-1000  ton
hydraulic  actuators,  which  act  as  four legs  of  a  table.  These
actuators are  inverted with their  pistons pushing against  the floor
rail plates  and their  upper bases pushing  against the  main support
carriage. Between the pistons and the floor rail plates are electrical
isolator compression pads. Electrical  isolators are also used between
the Hillman  Rollers and the floor  rail plates, once  the detector is
raised to  electrically isolate the detector from  the building. There
are mechanical locking collars on  each piston to support the detector
load when hydraulic power is removed. Both the horizontal and vertical
actuator  motions are  controlled  using a  closed  loop servo  system
allowing for  precise control of motion parameters  of both individual
or combined actuator systems.

The detector is  restrained from motion in both  the Assembly and Wide
Angle Halls by  the seismic supports.  These 4  large devices restrain
the detector in the horizontal plane, while maintaining the electrical
isolation preventing  multipoint grounding.  The  seismic supports are
capable  of   constraining  the   detector  from  0.25   g  horizontal
accelerations.

The 2 -  100 ton pole tips on  the East and West ends  of the detector
form the  end pieces  of the cylindrical  magnet steel.  They  must be
removable  for  accessing  the  internal detector  elements.  This  is
accomplished using a pole tip support carriage for each pole tip. Each
carriage weighs  70 tons and is  capable of lifting  and retracting or
inserting the  pole tip through  vertical, tilt, and  linear hydraulic
actuator motion. Once the pole tips are inserted, the carriages can be
disconnected and retracted away from the detector.

\section{Performance}

The design of STAR took many  years to complete.  Tools such as 3D CAD
were used to  simulate the detector and the  assembly.  Even with this
very careful planning process, it  was not possible to foresee all the
ramifications of  our decisions.  It certainly could  be possible that
an  overlooked  detail could  create  a  fundamental  conflict in  the
assembly of the detector.  Furthermore, we had to carefully manage any
last  minute  changes. These  items  could  introduce  delay into  the
project and increase the cost.  Furthermore, an unmanaged change could
negate the  planning processes.   The following sections  describe our
experience as the detector was assembled and subsequently operated.

\subsection{Assembly}

The  assembly  of  all  detector  elements went  very  smoothly.   The
sub-systems were monitored in reviews  to make sure that the designers
understood the envelope drawings  and interfaces.  Any change in their
design was carefully  reviewed to make sure that  the ramifications of
the modification  were understood and changes to  other detectors were
minimized.

There was no  need to modify or retrofit any part  of any detector due
to interference with  a neighboring element.  Nor was  there a need to
change  any items due  to mismatch  at an  interface.  There  was also
adequate  room for  all power  and signal  cables, gas  pipes, cooling
lines and  structural support elements.  Up front  thinking and system
level   design    (envelope   and   interface    drawings,   interface
specifications) of the STAR detector proved very worthwhile.

\subsection{Movement of the Detector with the Festoon}

The detector typically moves between the WAH and AH once per year. The
operation of stepping the detector the 33 m from one hall to the other
requires  4 hours.   Setup  and safety  checks  require a  significant
fraction of this time.

The festoon  lets us rapidly operate  the detector after  it is moved.
When the  festoon was first assembled,  the only copper  cables were a
safety/interlock  and  a  trigger  bundle cable.   After  the  initial
operation, the  copper trigger cable was disconnected  and replaced by
an optical fiber  connection.  As a 1 cm optical  fiber carries all of
the data from  one of the 24  TPC sectors, the size of  the festoon is
small. The  total diameter of  the full cable  bundle is about  15 cm.
This  amount includes  all of  the TPC,  EMC, SVT,  FTPC,  trigger and
miscellaneous  cables.  This  size  is very  small  compared to  cable
bundles that previous collider detectors needed.

When access is needed to the inner detector, it is necessary to remove
the  pole tips.  The  time to  disassemble both  the magnet  water and
electrical connections  and then move the  pole tip out of  the way is
about 8  hours.  The time to restore  the pole tip so  that the Magnet
can be operated is approximately the same time.

\subsection{Grounding} 

As  the  detector was  built,  it often  became  hard  to balance  the
requirements of low capacitance  with the physical requirements of the
detector.  Also, the  capacitance of the detector to  the walls of the
building was neglected in the design.

In the early  stages of the design when just the  Magnet and North and
South Platforms were attached, the  total capacitance was about 14 nf.
This number is significantly higher than the planned number of about 5
nf.  After the detector was fully assembled, the final value was about
130 nf.  First, this number  was measured with the pole tips retracted
from the magnet.   This particular stray capacitance is  not seen when
the detector is  taking data.  Secondly, a number  of sub-systems were
not  installed  as  planned.  For  instance, several  gas  pipes  were
installed very close to the North Platform, and several cable runs ran
on the concrete floor.

When the detector was installed, the GID showed that there was in fact
a stray path to ground.  This path vanished when the detector traveled
to the Assembly  Hall and the fault came  back intermittently when the
detector returned to the WAH.  After much investigation, a small metal
chip was found which shorted across a detector insulator.
  
The GID  is designed to measure  a ground fault of  0.17 $\Omega$.  At
one  time, we tried  to measure  the capacitance  of the  detector and
found that there was a 400 $\Omega$ ground fault.  Such a large number
could  not  be  measured with  the  GID  and  could  only be  done  by
disconnecting all  ground cables.  Similarly, the GID  is not designed
to measure  the capacitive  coupling.  A careful  measure of  the true
impedance coupling of the detector to the environment can only be done
when the ground connection is  disconnected.  This is not practical to
do often, as this measurement requires  all power to be turned off the
detector.

Tests show there is no significant external electrical contribution to
the  detector  noise from  the  surroundings  for  the TPC  pedestals.
Measurements made  show that  even with the  magnet operating  at full
current,  the pedestals  are  about  1.1 ADC  counts.   This value  is
approximately the same as measured in the laboratory.

\subsection{Rack Cooling}

The rack cooling supplies need  to hold the interior rack temperatures
low  enough  to  minimize  electronic  aging.  The  water  system  was
originally  designed   to  operate  at  16  C.    However,  the  water
temperature had to be increased to 19 C so that condensation would not
happen inside  the racks.  The  higher temperature was  needed because
the air conditioning system  could not dehumidify the air sufficiently
when  the outside  temperature was  hot and  the outside  humidity was
high.
    
Even though the water system ran at a much higher temperature, the VME
crate temperatures typically operate at  23 C, about 2 C above ambient
temperature and a very  comfortable temperature for electronics.  Even
the highest temperature in a crate was only 27 C.
 
Several  factors  contributed  to   this  good  fortune.   First,  the
preliminary estimates of  the power loads in the  crates turned out to
be  high.  For  a DAQ  crate, the  final value  was  approximately 700
w/crate, which  is about 1/2 of the  original projection. Furthermore,
the flow  for the cooling water was  specified to be 8  l/min for each
radiator.  The cooling system exceeded these specifications.  The flow
in the  DAQ heat  exchangers varied  between 12 to  18 l/min  with the
highest flow value at the bottom of the rack.

We could have  operated the input water temperature  at an even higher
water temperature.   However, the temperature of the  TPC water system
was coupled to  the rack water system and  any higher temperature made
the TPC  cooling system unstable.   Consequently, on the  most extreme
hot and humid  days, there was condensation in  the crates.  When this
condition happened,  the leak  detection system promptly  detected the
moisture and turned off the power  so that the moisture did not damage
the electronics.  A new dehumidifier  system has been added to the WAH
air conditioning system to eliminate this condition for the next run.

\section{Conclusion}

The team of integration and  system level engineers and physicists was
crucial to building the detector  on time and on budget.  The planning
worked  well and  as  a  result the  final  mechanical and  electrical
environment for STAR was built as intended.

\ack

We  would  like to  thank  the  engineering  and technical  staffs  of
Brookhaven   National  Laboratory   and  Lawrence   Berkeley  National
Laboratory for  their effort in building  STAR. We also  would like to
thank  the  RHIC  operations  group  for  their  support.  Doug  Fritz
contributed  to   mechanical  integration  for   this  project.   John
Scheblein, Bill Leonhard, Michael Cherney, and Tonko Ljubicic provided
us with vital information for  this paper.  Howard Wieman has reviewed
this  manuscript  and   contributed  several  thoughtful  suggestions.
Finally, we acknowledge the financial support from the U.S. Department
of Energy, Nuclear Physics.

\end{document}